\documentclass[aps,preprint]{revtex4}
\usepackage[dvips]{graphics,graphicx}

\begin{document}
\title{Landau-Zener tunneling in 2D periodic structures in the presence of a gauge field II: Electric breakdown}
\author{D.~N.~Maksimov$^1$, I.~Yu.~Chesnokov$^2$, D.~V.~Makarov$^3$, and A.~R.~Kolovsky$^{1,2}$}
\affiliation{$^1$Kirensky Institute of Physics, 660036 Krasnoyarsk, Russia}
\affiliation{$^2$Siberian Federal University, 660041 Krasnoyarsk, Russia}
\affiliation{$^3$Il'ichev Pacific Oceanological Institute, 690041 Vladivostok, Russia}
\date{\today}

\begin{abstract}
We analyze dynamics of a quantum particle in a square lattice in the Hall configuration beyond  the single-band approximation. For vanishing gauge (magnetic) field this dynamics is defined by the inter-band Landau-Zener tunneling, which is responsible for the phenomenon known as the electric breakdown. We show that in the presence of a gauge field this phenomenon is absent, at least, in its common sense. Instead, the Landau-Zener tunneling leads to appearance of a finite current which flows in the direction orthogonal to the vector of a potential (electric) field.
\end{abstract}
\maketitle

\section{Introduction}
\label{sec1}

This work continues our studies of the Landau-Zener tunneling (LZ-tunneling) for a quantum particle in the Hall configuration \cite{part1}. To be certain, here we assume a charged particle in electric and magnetic fields, although the results are equally applied to a neutral particle (for example, an atom in an optical lattice) subject to artificial gauge and potential fields. Specifically, we consider the following system 
\begin{equation}
\label{1}
\widehat{H}=\frac{1}{2M}\left[\hat{p}_x^2 + (\hat{p}_y - \frac{e}{c}B x)^2\right] + V(x,y)+e(F_x x+F_y y) \;,
\end{equation}
where $B$ is the magnitude of a magnetic field,  ${\bf F}=(F_x,F_y)$ an electric field, and $V(x,y)=v_x\cos (2\pi x/a)+v_y\cos(2\pi y/a)$ the periodic potential. For $B=0$ the spectrum of the system (\ref{1}) consists of Bloch bands separated by energy gaps. It is well know that a strong electric field induces Landau-Zener transitions (tunneling) across the energy gaps. In our previous work \cite{part1} we have shown that these transitions also take place in the presence of a magnetic field and calculated the tunneling rate. Unfortunately, the analytical approach of Ref.~\cite{part1} provides only the rate of tunneling but is incapable to describe dynamics of the system (\ref{1}) if LZ-tunneling is not negligible. In the present work we study dynamical manifestations of LZ-tunneling by numerically solving the Schr\"odinger equation with the Hamiltonian (\ref{1}) and comparing the results with those obtained on the basis of the tight-binding model (tb-model),   
\begin{equation}
\label{2}
\left(\widehat{H}_{tb}\Psi\right)_{l,m}=  
 -\frac{J_x}{2} \left( \psi_{l+1,m}+ \psi_{l-1,m} \right)
-\frac{J_y}{2}  \left( \psi_{l,m+1}e^{i 2\pi\alpha l} +  \psi_{l,m-1}e^{-i 2\pi\alpha l} \right)  
+ea (F_x l +F_y m) \psi_{l,m} \;,
\end{equation}
which is believed to approximate the Hamiltonian (\ref{1}) in the case of negligible LZ-tunneling. In Eq.~(\ref{2})   $J_{x,y}$ are the hopping matrix elements in two orthogonal directions, $a$ is the lattice period, $l=x/a$ and $m=y/a$ label lattice sites, and $\alpha$ is the Peierls phase,
\begin{equation}
\label{3}
\alpha=\frac{eBa^2}{hc} \;.
\end{equation}
It should be mentioned that the tight-binding approximation implicitly  assumes a deep lattice. For the considered periodic potential this requires $v_{x,y}\ge E_R/2$, where $E_R=\hbar^2/Ma^2$. If the later condition is satisfied the ground Bloch band of the one-dimentional Hamiltonians $\widehat{H}_0^{(x,y)}$,
\begin{equation}
\label{4}
\widehat{H}_0^{(x)}=\frac{\hat{p}_x^2}{2M} - v_x\cos\left(2\pi\frac{x}{a}\right)  \;,
\end{equation}
($\widehat{H}_0^{(y)}$ is given by the similar equation) are well approximated by the cosine function, $E^{(x)}(\kappa_x)=E_0^{(x)}-J_x\cos(a\kappa_x)$. Thus the next to neighboring hopping terms can be indeed neglected in the Hamiltonian (\ref{2}). 
In what follows we measure the energy in units of $E_R$, the length in units of $a$ (which we set to $2\pi$), and drop all physical constants. 
\begin{figure}[t!]
\center
\includegraphics[width=7cm, clip]{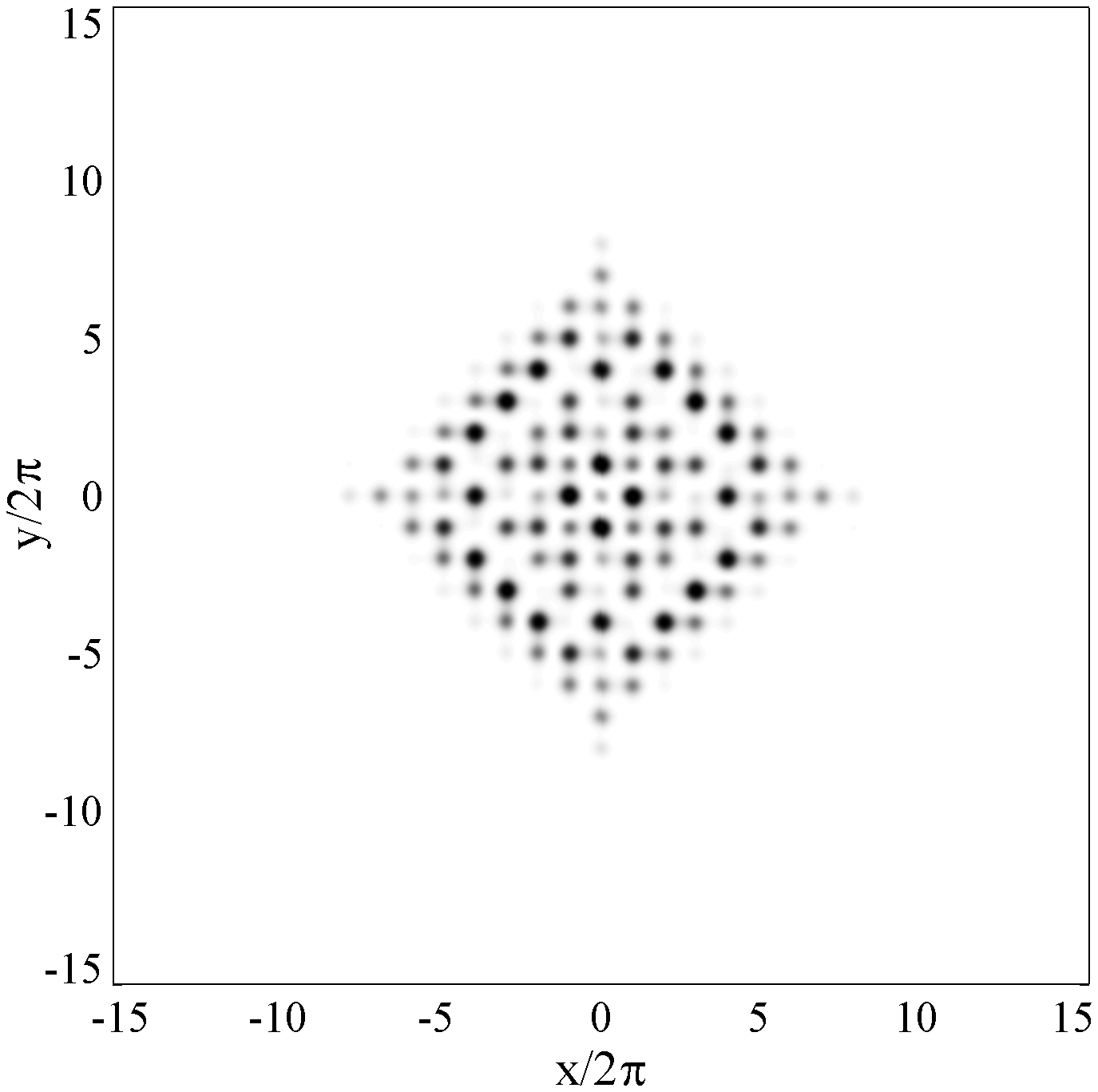}
\includegraphics[width=7.2cm, clip]{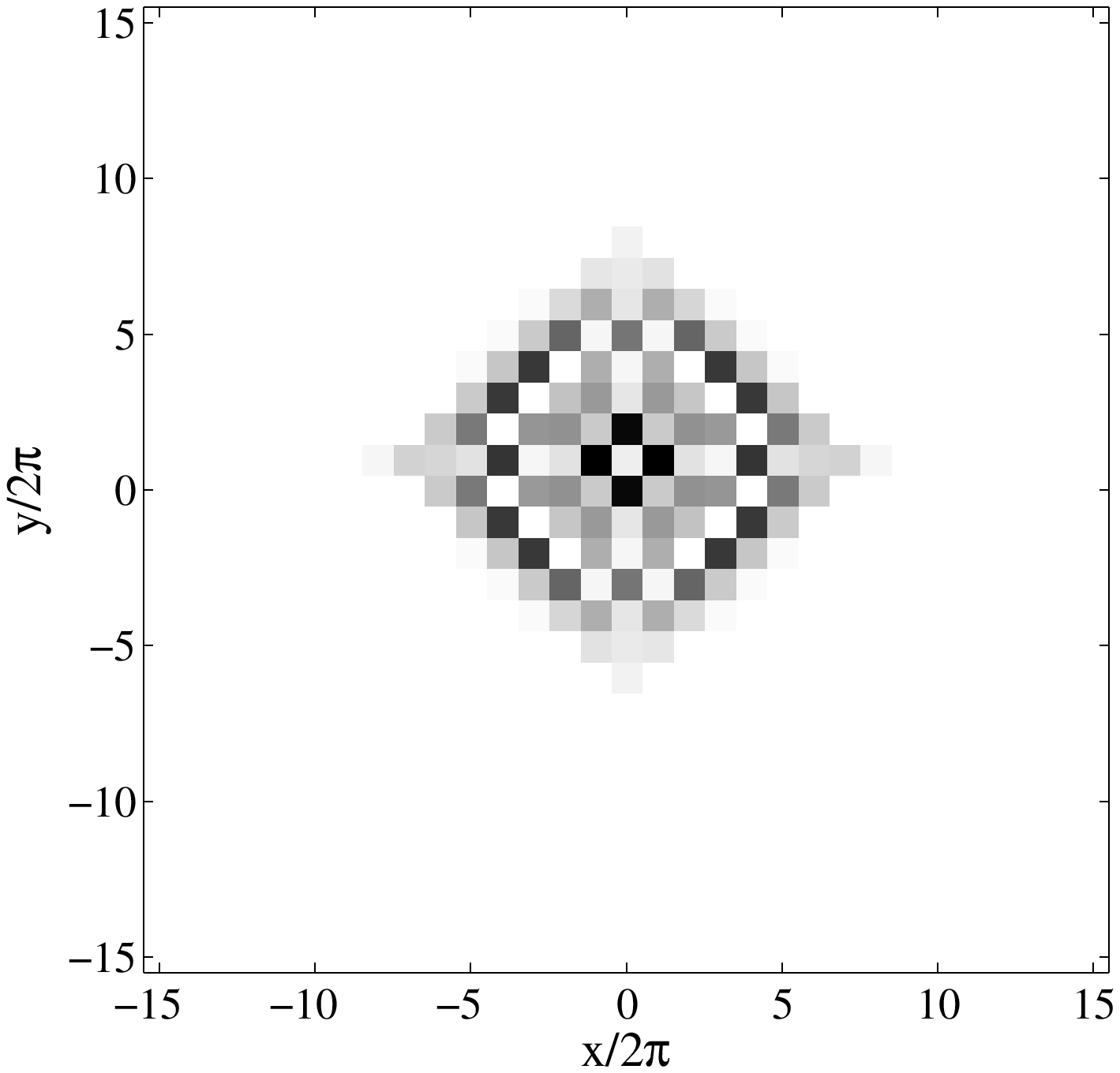}
\caption{Squared absolute value of the continuous wave-function $\Psi(x,y,t)$, left, and squared absolute values of the amplitudes $\psi_{l,m}(t)$, right, for $t=300$. The system parameters are $v_x=v_y=0.5$, $F=0$,  and $B/2\pi=\alpha=1/8$.}   
\label{fig4}
\end{figure}

\section{Vanishing electric field}
\label{sec3}

We begin with comparing dynamics of the systems (\ref{1}) and (\ref{2}) for $F=0$. We solve the time-dependent Schr\"odinger equation with the Hamiltonian (\ref{1}) on the grid and check the convergence of the results with respect to the space and time discretization.  Having the wave-function obtained, we calculate its overlap with the Wannier functions $\Phi_{l,m}(x,y)$ associated with the ground Bloch band of the Hamiltonian $\widehat{H}_0= \widehat{H}_0^{(x)} + \widehat{H}_0^{(y)}$:
\begin{equation}
\label{b1}
\psi_{l,m}(t)=\int {\rm d}x{\rm d}y \Phi_{l,m}(x,y) \Psi(x,y,t) \;,\quad 
\Phi_{l,m}(x,y)=\Phi_{0,0}(x-2\pi l,y-2\pi m) \;.
\end{equation}
Finally the probability amplitudes (\ref{b1}) are compared with those calculated on the basis of the tight-binding Hamiltonian (\ref{2}).

A remark concerning initial conditions is in order. To facilitate comparison with tb-model we choose $\Psi(x,y,t=0)=\Phi_{0,0}(x,y)$,  that corresponds to population of the single site, $\psi_{l,m}(t=0)=\delta_{l,0}\delta_{m,0}$. Notice that this initial state assumes population of the whole ground Bloch band if $B=0$ or all ground magnetic bands  if $B\ne0$ \cite{part1}.  As an example the left panel in Fig.~\ref{fig4} shows the squared absolute value of the continuous wave-function $\Psi(x,y,t)$ for the specified initial condition after evolution time of the order of one cyclotron period $T_c=2\pi/\omega_c$ \cite{remark4}. The right panel in Fig.~\ref{fig4} shows the squared absolute values of the amplitudes (\ref{b1}). For the considered evolution time the depicted in the right panel populations of the lattice sites are undistinguishable by eye from those obtained on the basis of tb-model. However, for larger times we observe clear deviations between the  original system and tb-model, which we discuss in the next paragraph.
\begin{figure}[t]
\center
\includegraphics[width=11.5cm, clip]{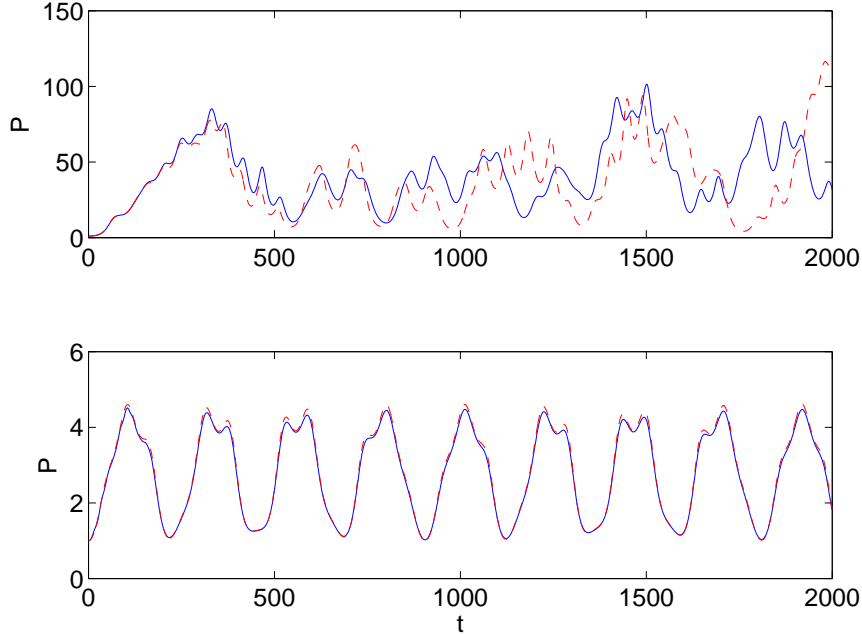}
\caption{Upper panel: participation ratio (\ref{b3}) as functions of time for the parameters of Fig.~\ref{fig4}. Lower panel: the same yet $F=0.015$,  $F_x/F_y=(\sqrt{5}-1)/4\approx 0.309$.  Dashed lines show the participation ratio calculated by using tb-model.}   
\label{fig5}
\end{figure}
\begin{figure}
\center
\includegraphics[width=10cm, clip]{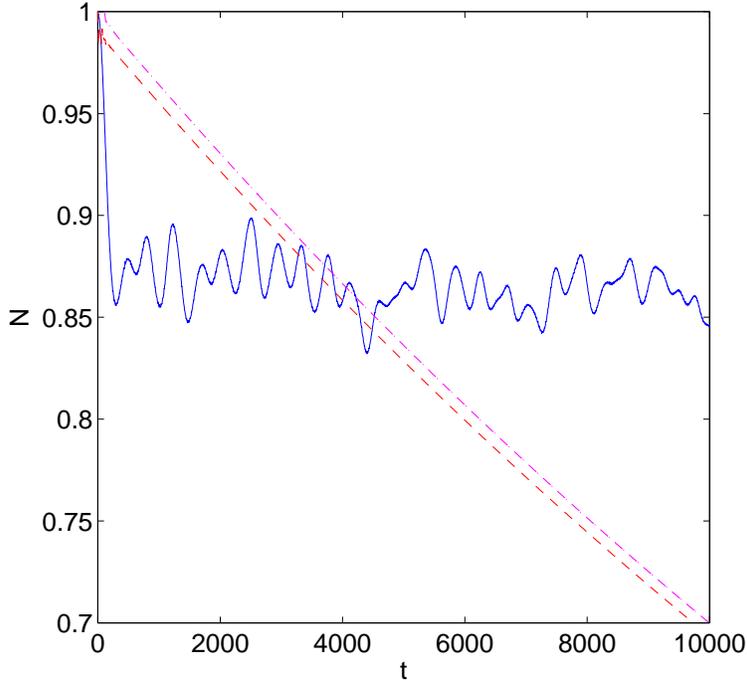}
\caption{The projection (\ref{b4}) on the subspace spanned by the ground WS-states for $F=0$ (solid line) and $F=0.015$ (dashed line). The dash-dotted line shows the decrease of total probability in the case $F=0.015$ due to absorbing boundary conditions.}   
\label{fig10}
\end{figure}

As an overall characteristic of the system dynamics we consider the participation ratio
\begin{equation}
\label{b3}
P(t)=\left(\sum_{l,m} |\psi_{l,m}(t)|^4 \right)^{-1} \;.
\end{equation}
The dashed line in the upper panel in Fig.~\ref{fig5} depicts $P(t)$ calculated by using tb-model. The participation ratio shows oscillatory dynamics,  superimposed with a slow increase in its mean value. This slow increase is due to finite widths of the magnetic bands, which implies ballistic dynamics for asymptotic times \cite{remark5}. The dashed line should be compared with the solid line, which is calculated by using the original Hamiltonian. Although qualitative correspondence is noticed, quantitative agreement holds only for finite time.  The reason for the observed deviations is a relatively large value of $\alpha$ used in our numerical simulations. In fact, if $B\ne0$ the condition $v_{x,y}\ge 1/2$ alone is not enough to justify the tight-binding Hamiltonian. The other necessary condition, which is often overlooked when deriving (\ref{2}) from the continuous Hamiltonian (\ref{1}), is $\alpha\ll 1$. Indeed, if $\alpha\ne0$ dynamics of the system (\ref{1}) is not restricted to the subspace of the Hilbert space spanned by the ground Wannier states $\Phi_{l,m}(x,y)$, even if the initial wave-function belongs to this subspace. This statement is illustrated in Fig.~\ref{fig10} which shows the projection of the wave function on the specified subspace,
\begin{equation}
\label{b4}
N(t)=\sum_{l,m} |\psi_{l,m}(t)|^2 \;.
\end{equation}
Obviously, the smaller $\alpha=2\pi B$, the closer $N(t)$ is to unity. However, smaller $\alpha$ require larger lattices and longer evolution time to see characteristic features of the system dynamics. Our choice $\alpha=1/8$ is a compromise between accuracy of the tight-binding approximation and complexity (computational time) of numerical simulations.

\section{Vanishing magnetic field}
\label{sec4}

\begin{figure}
\center
\includegraphics[width=10cm, clip]{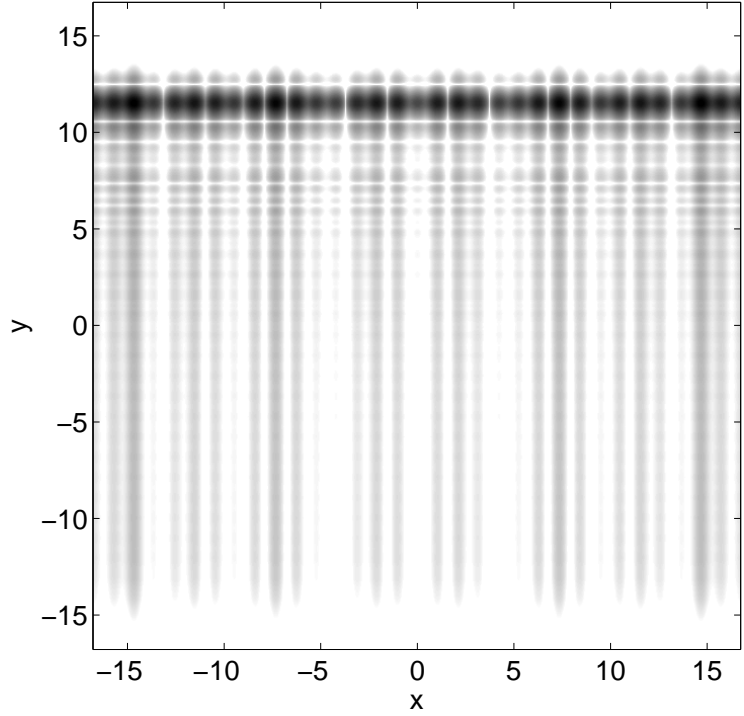}
\caption{Logarithm of  $|\Psi(x,y,t)|^2$ at  $t=10000$ for vanishing magnetic field. The system parameters are $v_x=v_y=0.5$, $F_y=0.015$, and $F_x=0$.}   
\label{fig6}
\end{figure}

Next we consider the case $F\ne0$ and $B=0$. For $B=0$ dynamics of the system (\ref{1}) are Bloch oscillations, where the mean position (normal mode) or the width (breathing mode) of a localized wave packet oscillates with the Bloch frequencies 
\begin{equation}
\label{b5}
\omega_B^{(x,y)}=eaF_{x,y}/\hbar=2\pi F_{x,y} \;.
\end{equation}
%
Simultaneously, the oscillating  wave packet emits sub-packets, which get accelerated by the electric field and move away from the main packet along the crystallographic axes of the lattice \cite{55,58}. This dynamics is exemplified  in Fig.~\ref{fig6}, which shows $|\Psi(x,y,t)|^2$ for $F_y=0.015$ and $F_x=0$, after evolution time $t=10000$. For the chosen electric field configuration the sub-packets move in the negative $y$ direction while the packet itself spreads in the $x$ direction. Since in numerical simulations we are forced to deal with finite lattices, we impose periodic boundary conditions at $x=\pm 16$ and absorbing boundary conditions at $y=-16$. For this setup the system quickly reaches quasi stationary state, where time-average occupation probabilities of the lattice sites do not depend on $l$. Notice that due to absorbing boundary condition the total probability decay in time. The dash-dotted and dashed lines in  Fig.~\ref{fig10} show the total probability and the projection (\ref{b4}) on the subspace spanned by the ground WS-states. These two curves are seen to follow each other. Thus, unlike the case of null electric field, deviations from tb-model are now exclusively due to irreversible LZ-tunneling to higher bands. 

Along with the quantity (\ref{b4}) the other important characteristics of the system is the mean kinetic energy $E_K(t)$ of the particle. Since the total energy is conserved, an increase in the kinetic energy is compensated by decrease in the Stark energy
\begin{equation}
\label{b6}
E_S(t)=\int {\rm d}x{\rm d}y |\Psi(x,y,t)|^2(F_x x + F_y y) \;.
\end{equation}
It is easy to prove that $E_K\approx-E_S(t)$ infinitely grows due to LZ-tunneling. This is in strong contrast with the case $B\ne0$, which we shall study systematically in the next section. For the moment we only mention that a finite $B$ prohibits infinite increase/decrease in the kinetic/Stark energy and imposes an upper boundary for possible energies of the quantum particle. This boundary can be estimated by using the classical arguments as follows.
\begin{figure}
\center
\includegraphics[width=10cm, clip]{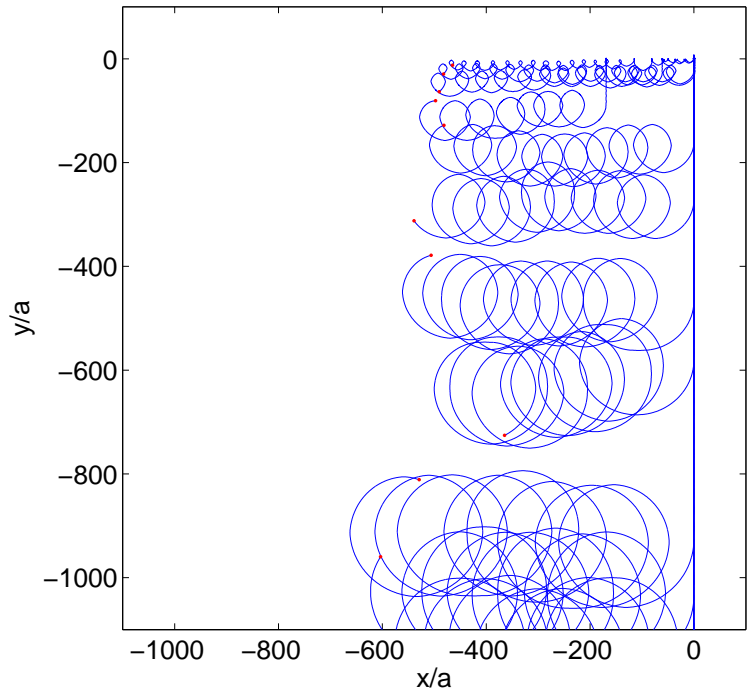}
\caption{Classical trajectories of 10 particles with random initial conditions in the central potential. Parameters are $v_x=v_y=0.5$, $B=0.02$, $F_y=0.02$, and $F_x=0$.}   
\label{fig7}
\end{figure}

Let us consider classical counterpart of the Hamiltonian (\ref{1}) where, for the sake of simplicity, we assume $F_x=0$. Let us also assume that the particle is initially located at $(x,y)\approx 0$ and its energy is slightly higher than the potential barrier in the $y$ direction. Then, for finite time defined below, the particle is accelerated by the electric field in the $y$ direction and $p_y(t)\approx Ft$. During this initial stage the motion in the $x$ direction is oscillations around the central local minimum  of the effective potential $V(x,t)=-v_x\cos x+(Ft-Bx)^2/2$. It is easy to see that this local minimum disappears if $t>t^*$, where 
\begin{equation}
\label{b7}
t^*=\frac{v_x+B^2 \pi/2}{BF} \approx \frac{v_x}{BF}  \;.
\end{equation}
At this time the particle escapes the central valley of the 2D potential with the kinetic energy 
\begin{equation}
\label{b8}
E_{max}\approx (Ft^*)^2/2 \sim (v_x/B)^2 \;.
\end{equation}
As soon as the particle escapes the valley its further trajectory becomes usual cycloid, see Fig.~\ref{fig7}, where the kinetic and Stark energies oscillates around $E_{max}$. The above analysis can be easily generalized for other directions of the electric field vector, resulting in the same estimate. It should be also mentioned that Eq.~(\ref{b7}) gives the maximal escape time. If we consider an ensemble of particles with slightly different initial conditions, we get different escape times and, as the consequence, a broad distribution for the kinetic/Stark energies.

\section{Finite magnetic and electric fields}
\label{sec5}

\begin{figure}
\center
\includegraphics[width=10cm, clip]{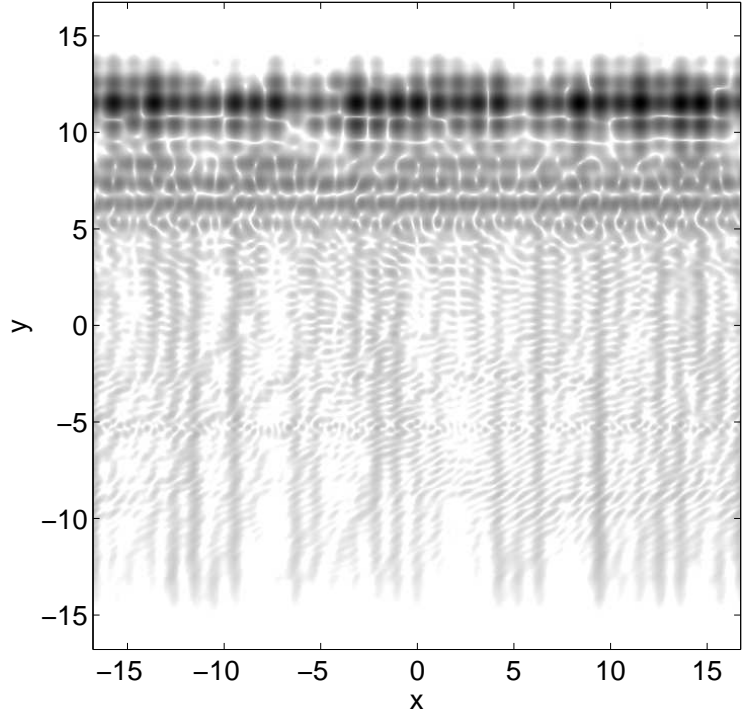}
\caption{Logarithm of $|\Psi(x,y,t)|^2$ at $t=10000$. Parameters are $v_x=v_y=0.5$, $\alpha=1/8$ ($B\approx0.02$), $F_y=0.015$, and $F_x=0$.}   
\label{fig8}
\end{figure}

We proceed with the case of $F\ne0$ and $B\ne0$. First we consider the situation where ${\bf F}$ is aligned with the $y$ axis.  Figure \ref{fig8} shows $|\Psi(x,y,t)|^2$ for $F=0.015$ and $\alpha=1/8$ ($B\approx0.02$) and should be compared  with Fig.~\ref{fig6} where $B=0$. Qualitative difference between two figures is that in the former case the tunneling particle acquires finite velocity in the $x$ direction, which is consistent with the classical dynamics considered in Sec.~\ref{sec4}. We calculated the mean current along the $x$ direction, which was found to differ from zero. Notice that tb-model predicts zero current for the considered initial condition. Thus a finite current in the $x$ direction can be considered a manifestation of LZ-tunneling.

\begin{figure}
\center
\includegraphics[width=9cm, clip]{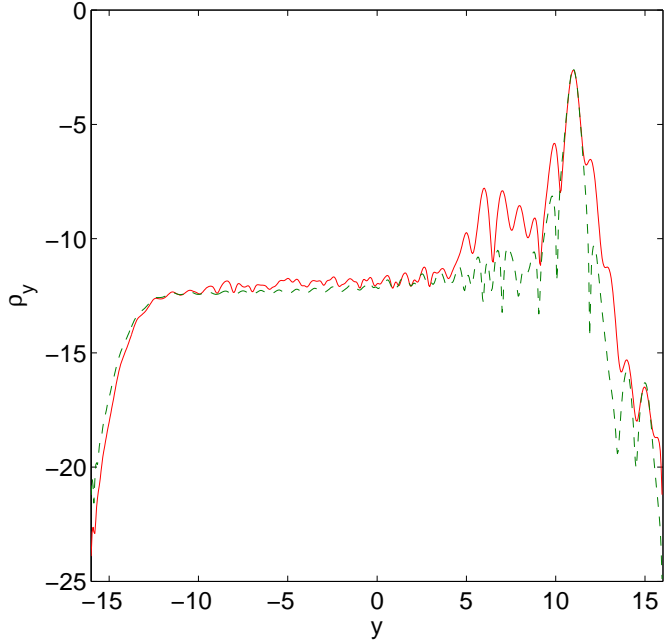}
\caption{Reduced probability $\rho(y)$ at $t=10000$ in the logarithmic scale. The solid and dashed lines correspond to $\alpha=1/8$ and $\alpha=0$, respectively.}   
\label{fig9}
\end{figure}

Unfortunately, we have not been able to find the actual value of the current. To do this one should move the absorbing boundary further to negative $y$ till the convergence is reached. [According to the estimate (\ref{b7}) this requires the lattice of the order of 1000 sites, which is beyond our computational capabilities.]  In this case we also expect to see a qualitative  difference in the tails of the reduced probability
\begin{equation}
\label{c1}
\rho(y)=\int  |\Psi(x,y,t)|^2 {\rm d}x  \;,
\end{equation}
see Fig.~\ref{fig9}, which defines the Stark energy (\ref{b6}) trough the equation $E_S=F\int y\rho(y){\rm d}y$. Based on the classical consideration, if $B\ne0$ the reduced probability is expected to decay faster than $1/|y|$ which ensures finite kinetic and Stark energies. On the contrary, if $B=0$ the reduced probability should show a diverging tail because the Stark and kinetic energies infinitely  increase when $t\rightarrow\infty$.  

\begin{figure}
\center
\includegraphics[width=7cm, clip]{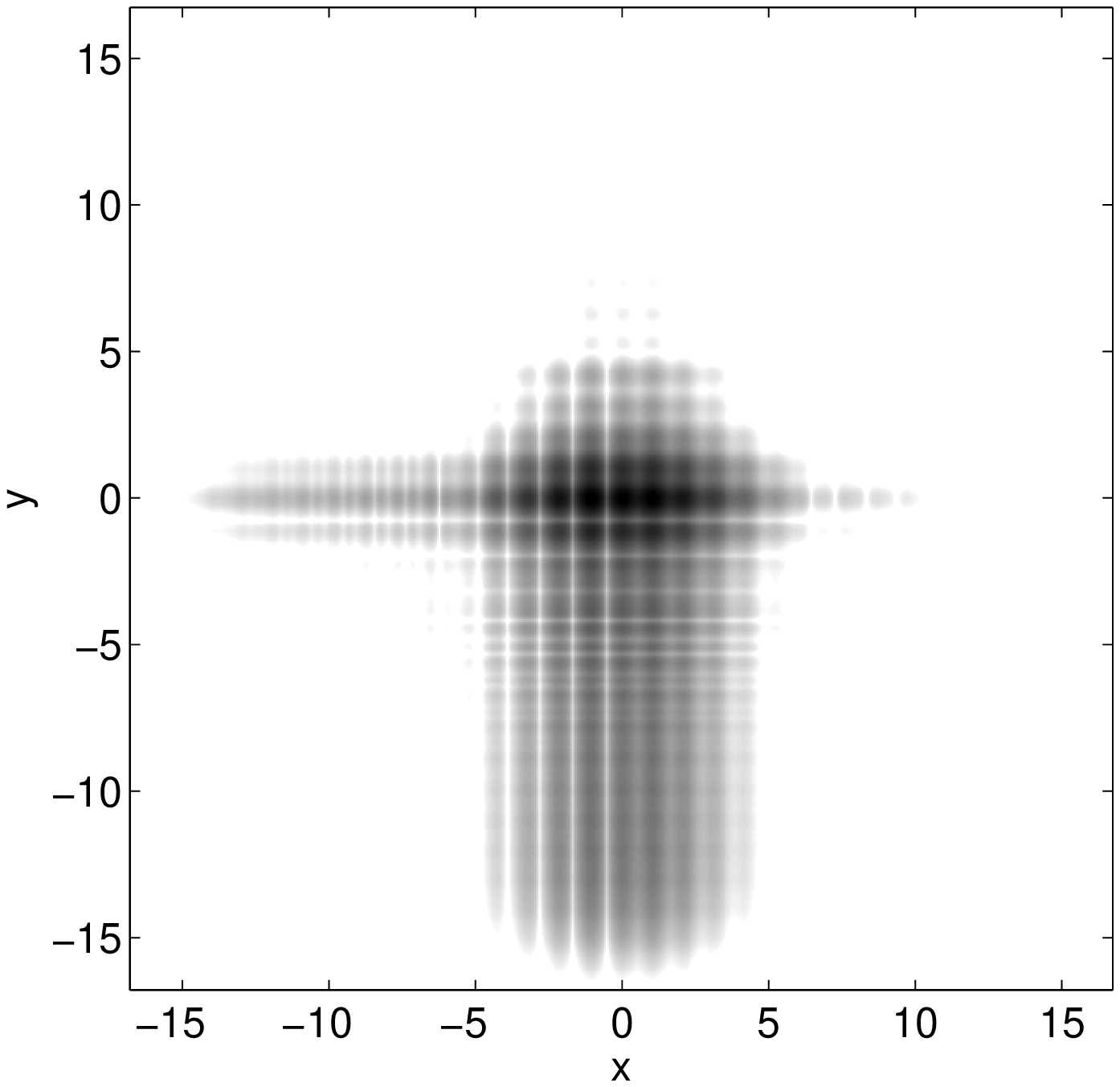}
\includegraphics[width=7cm, clip]{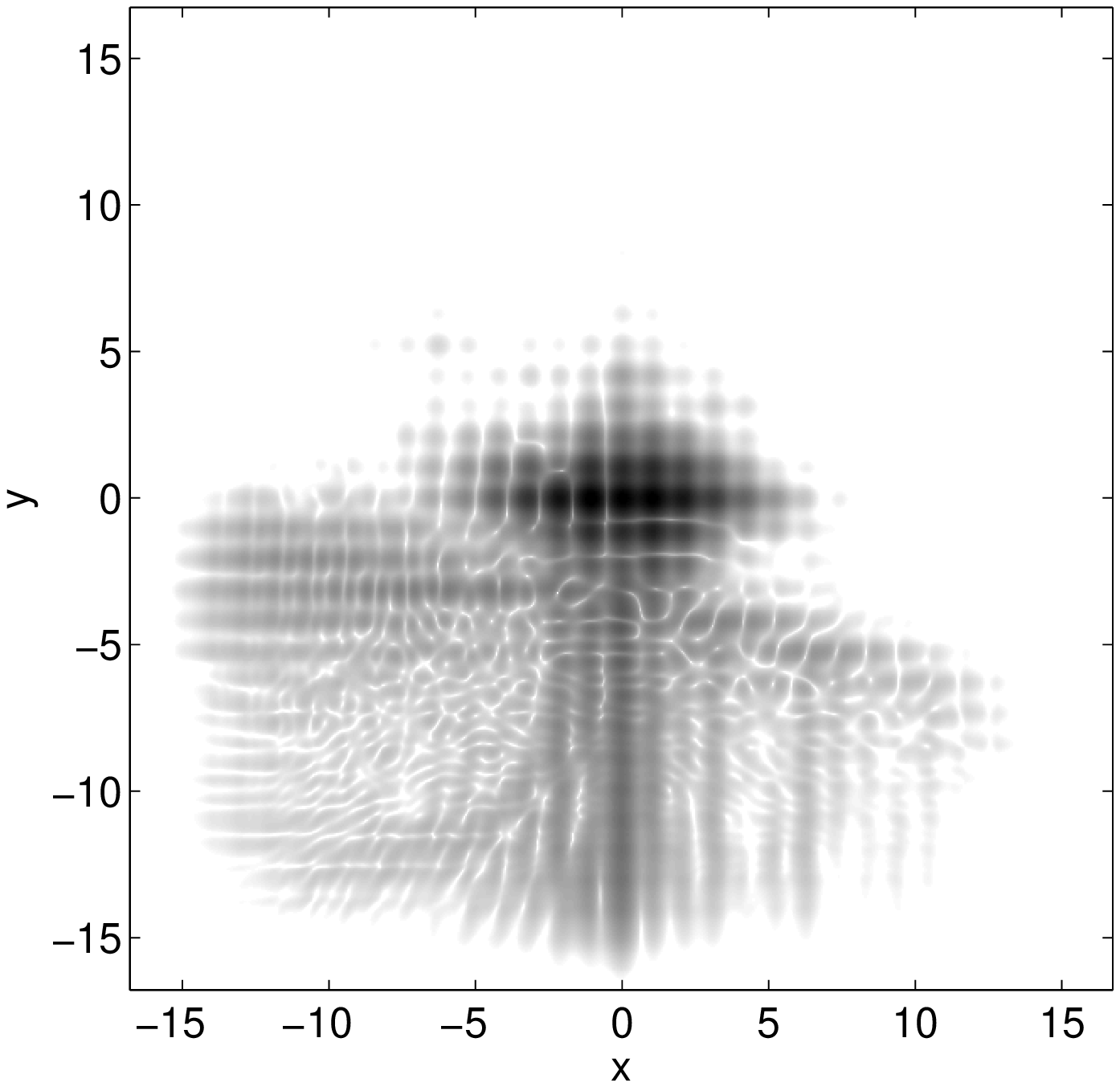}
\caption{Logarithm of $|\Psi(x,y,t)|^2$ at $t=10000$ for $\alpha=0$ (left panel) and $\alpha=1/8$ (right panel). The other parameters are $v_x=v_y=0.5$,  $F=0.015$, and $F_x/F_y=(\sqrt{5}-1)/4$.}   
\label{fig11}
\end{figure}

Finally we consider the case of an irrational direction of the electric field $F_x/F_y=(\sqrt{5}-1)/4\approx 0.309$. 
As shown in the recent works \cite{85,90,91} devoted to dynamical and spectral properties of the system (\ref{2}), in this case the electric field localizes the particle in the lattice, where the localization length tends to one site for $F/J \gg1$.  Notice that this result holds for both zero and finite magnetic fields. Numerical simulations of the original system for $F=0.015$ partially confirm this conclusion: When plotting the wave-packet in the linear scale we observe that only 3-4 sites participates in the dynamics. Moreover, if we normalize the amplitudes (\ref{b1}) against $N(t)$ the participation ratio (\ref{b3}) practically coincides with that calculated on the basis of tb-model, see lower panel in Fig.~\ref{fig5}.  However, when plotting the wave packet in the logarithmic scale we clearly see the effect of LZ-tunneling, as well as the qualitative difference between the cases $B=0$ and $B\ne0$, see Fig.~\ref{fig11}. 

\section{Conclusions}

We analyzed LZ-tunneling for a charged particle in a square lattice subjected to normal to the lattice plane magnetic field $B$ and in-plane electric field $F$. For $B=0$ the rate of LZ-tunneling is defined by the size of the  energy gap $\Delta$ separating the ground Bloch band from the rest of the spectrum and the magnitude of the electric field. If the rate of this tunneling is not negligible (strong electric fields), the ground Bloch band rapidly depletes and the particle is accelerated towards infinite kinetic energy -- the phenomenon known as electric breakdown.  We showed that a finite magnetic field essentially modifies this phenomenon, where the effect of the magnetic field is twofold. 

First, the magnetic field splits Bloch bands into magnetic bands, thus introducing the new energy gaps. According to results of our previous work \cite{part1}, now the rate of tunneling across the main energy gap $\Delta$ is either smaller or larger than that for $B=0$, depending on which magnetic band is initially populated. However, as the first approximation, one can neglect this effect and use the standard Landau-Zener equation to estimate the rate of tunneling.  Numerical results presented in this work undoubtedly confirm this conclusion of Ref.~\cite{part1}.

The other effect of the magnetic field, which cannot be neglected, is that the tunneled particle is accelerated only to a finite kinetic energy, that is in strong contrast with the case $B=0$. Using classical arguments we estimated the upper boundary for the particle kinetic energy, see Eq.~(\ref{b8}). After the acceleration stage, the further dynamics of the classical particle is a drift  in the direction orthogonal to the vector ${\bf F}$, i.e., the Hall current.  Our numerical simulations of the quantum system clearly indicate the presence of this current. Thus, the electric breakdown in the presence of a magnetic field should be interpreted as appearance of the Hall current.

The authors acknowledge financial support of Russian Academy of Sciences through the SB RAS integration project  No.29 {\em Dynamics of atomic Bose-Einstein condensates in optical lattices} and the RFBR project No.12-02-00094 {\em Tunneling of the macroscopic quantum states}. D.V.M. also acknowledges the support of RFBR through grants No.12-02-90807 and No.12-02-31416.


\begin{thebibliography}{10}
\bibitem{remark4}
In dimensional variables the cyclotron frequency is given by  $\omega_c=eB/cM^*$ where  $M^*=\hbar^2/a^2\sqrt{J_x J_y}$ is the effective mass.
\bibitem{remark5}
Well-known exclusion is the case of irrational $\alpha$ and $J_x=J_y$, where the spectrum of the tight-binding Hamiltonian is singular continuous and the asymptotic dynamics is anomalous diffusion, also known as quantum intermittency \cite{Guar94}.
\end{thebibliography}
\end{document}